\documentclass[12pt]{revtex4}
\usepackage{amsfonts}
\usepackage{graphicx, amsmath}
\usepackage{longtable}
\makeatletter

\newcommand{\Rmnum}[1]{\expandafter\@slowromancap\romannumeral #1@}
\makeatother
\begin{document}
\title[Short Title]{Feasible superadiabatic-based shortcuts for fast generating 3D entanglement between two atoms}
\author{Xiao-Qin~Yang}
\author{Dian-Yang~Huang}
\author{Peng~Xue}
\author{Yong-Yong~Gong}
\author{Jin-Lei Wu}
\author{Xin Ji\footnote{E-mail: jixin@ybu.edu.cn}}
\affiliation{Department of Physics, College of Science, Yanbian University, Yanji, Jilin 133002, People's Republic of China}

\begin{abstract}
\noindent \textbf{{Abstract}} We propose a scheme to realize fast generation of three-dimensional entanglement between two atoms via superadiabatic-based shortcuts in an atom-cavity-fiber system. The scheme is experimentally feasible because of the same form of the counterdiabatic Hamiltonian as that of the effective Hamiltonian. Besides, numerical simulations are given to prove that the scheme is strongly robust against variations in various parameters and decoherence.
\end{abstract}
\maketitle
\section{Introduction}
With the rapid development in quantum information processing, high-dimensional entanglement is increasingly drawing attention of researchers due to its more superior security than qubit entanglement in the field of quantum key distribution and its greater violation of local realism~\cite{BKB2001,BM2002,CBK2002,YW2016,DPMW2000}. Thus, the generation of high-dimensional entanglement is of great importance. Up to present, a large number of schemes have been proposed for generating high-dimensional entanglement via various techniques~\cite{XHL2010,LP2012,YLS2015,SSW2016,WG2011,SJR2013,QCW2013,XQ201402,SL2014,DJC2015}. Among these techniques, stimulated Raman adiabatic passage~(STIRAP) is widely used in fields of time-dependent interaction for many purposes~\cite{KHB1998,PIM2007} because of its robustness against atomic spontaneous emissions and variations in experimental parameters. However, STIRAP usually requires a relatively long interaction time for restraining non-adiabatic transitions.

A set of techniques called ``Shortcuts to adiabaticity~(STA)" are promising for quantum information processing which actually fights against the decoherence, noise, or losses that are accumulated during a long operation time. Hence, many schemes are proposed to
construct STA~\cite{XCA2010,XASA2010,SIX2012,dC2012,ARXD2012,SMG2013,AC2013,ETS2013,DGO2014,YHC2016,AHA2016}. By using STA, a great deal of remarkable achievements have been made in quantum information processing~\cite{MYLJ2014,YHC2014,XL2015,JTD2015,YLQ2015,XHQ2016}. Also, numerous schemes have been come up with for fast generating high-dimensional entanglement~\cite{JYC2016,ZYY2016,WSJ2016,HSW2016,WJZ2016}, in which Chen $et~al$ and He $et~al$ prepared a three-atom singlet state~\cite{ZYY2016} and a two-atom 3D entangled state~\cite{HSW2016}, respectively, by using transitionless quantum driving~(TQD); Lin $et~al$~\cite{JYC2016} and Wu $et~al$~\cite{WSJ2016} implemented two-atom 3D entangled states, respectively, based on Lewis-Riesenfeld invariants~(LRI); Wu $et~al$ also generated three-atom tree-type 3D entangled states with both of TQD and LRI~\cite{WJZ2016}.

In this work, we propose a superadiabatic scheme for fast generating two-atom 3D entanglement via superadiabatic iterations. Superadiabatic iterations as an extension of the traditional adiabatic approximation was introduced in~\cite{BERRY1987}. The technique was adopted for speeding up adiabatic process first by Ib\'{a}\~{n}ez $et~al$~\cite{SIX2012,SXJ2013}. A short time before, Song $et~al$ extended it to a three-level system~\cite{XQJ2016}. More recently, Huang $et~al$~\cite{HCW2016} and Kang~\cite{YYQ2016} generated Greenberger--Horne--Zeilinger state and $W$ state, respectively, by using this technique. Now we apply this technique to the fast generation of two-atom 3D entanglement. Apart from the rapid rate, we implement two-atom 3D entanglement with pretty high fidelity. More importantly, as the second iteration different from the first iteration~(i.e., TQD), the superadiabatic scheme does not need an additional coupling between the initial and finial states, and the same form of counterdiabatic Hamiltonian as that of effective Hamiltonian guarantees its high feasibility in experiment.

This paper is structured as follows: The physical model and effective dynamics are shown in section~2. In section 3, we give the superadiabatic scheme for fast generating 3D entanglement between two atoms. In section~4, numerical simulation results prove that the scheme is fast, valid and robust. The conclusion is given in section~5.
\section{Physical model and effective dynamics}
\begin{figure}[htb]\centering
\centering
\includegraphics[width=\linewidth]{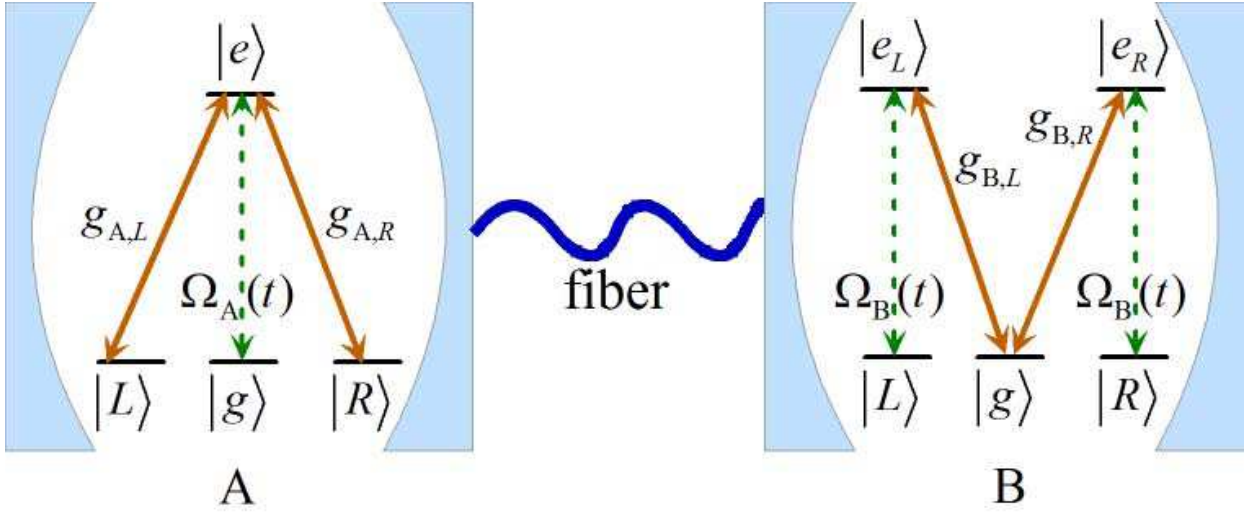}
\caption{The diagrammatic sketch of the atom-cavity-fiber system, atomic level configurations and related transitions.}\label{f1}
\end{figure}
The schematic sketch of the physical model for fast generating two-atom 3D entanglement is shown in figure~\ref{f1}. There are two atoms trapped, respectively, in two spatially separated bimodule cavities connected by a fiber which satisfies the short fiber limit to ensure that only resonant modes of the fiber interact with cavity modes~\cite{SMB2006}. Two atoms both have three ground states $|L\rangle$, $|g\rangle$ and $|R\rangle$. In addition, atom \rm A has one excited state $|e\rangle$ and atom \rm B has two excited states $|e_L\rangle$ and $|e_R\rangle$. Atomic transitions $|e\rangle_{\rm A}\leftrightarrow|L(R)\rangle_{\rm A}$ and $|e_{L(R)}\rangle_{\rm B}\leftrightarrow|g\rangle_{\rm B}$ are resonantly coupled to the left(right)-circularly polarized modes of cavity {\rm A} and cavity {\rm B}, respectively, with corresponding coupling constants $g_{{\rm A},{L(R)}}$ and $g_{{\rm B},{L(R)}}$. Transitions $|e\rangle_{\rm A}\leftrightarrow|g\rangle_{\rm A}$ and $|e_{L(R)}\rangle_{\rm B}\leftrightarrow|L(R)\rangle_{\rm B}$ are resonantly driven by classical laser fields, respectively, with Rabi frequencies $\Omega_{\rm A}(t)$ and $\Omega_{\rm B}(t)$. Then, the interaction Hamiltonian of the atom-cavity-fiber system is~($\hbar=1$):
\begin{eqnarray}\label{e1}
H(t)&=&\Omega_{\rm A}(t)|g\rangle_{\rm A}\langle{e}|+\sum_{i=L,R}[\Omega_{\rm B}(t)|i\rangle_{\rm B}\langle{e_i}|+g_{{\rm A},i}a_{{\rm A},i}|e\rangle_{\rm A}\langle{i}|\nonumber\\
&&+g_{{\rm B},i}a_{{\rm B},i}|e_i\rangle_{\rm B}\langle{g}|+v b_i(a_{{\rm A},i}^\dag+a_{{\rm B},i}^\dag)]
+\rm H.c.,
\end{eqnarray}
where $a_{{\rm A(B)},L(R)}$ and $b_{L(R)}$ is the annihilation operator of left(right)-circularly polarized mode of cavity \rm A(B) and the fiber, respectively; $v$ is the coupling strength between the two cavities and the fiber. For convenience, we assume $g_{{\rm A(B)},L(R)}$ is real, $g_{{\rm A},{L}}=g_{{\rm A},{R}}=g_{\rm A}$ and $g_{{\rm B},{L}}=g_{{\rm B},{R}}=g_{\rm B}$.

If the initial state of the whole system is $|\phi_1\rangle=|g\rangle_{\rm A}|g\rangle_{\rm B}|0\rangle_{c \rm A}|0\rangle_f|0\rangle_{c \rm B}$ denoting two atoms both in state $|g\rangle$ and two cavities and the fiber all in the vacuum state, Hamiltonian~(\ref{e1}) can be rewritten by
\begin{eqnarray}\label{e2}
H(t)&=&\Omega_{\rm A}(t)|\phi_1\rangle\langle\phi_2|+g_{\rm A}|\phi_2\rangle(\langle\phi_3|+\langle\phi_4|)+v(|\phi_3\rangle\langle\phi_5|+|\phi_4\rangle\langle\phi_6|+|\phi_5\rangle\langle\phi_7|+|\phi_6\rangle\langle\phi_8|)\nonumber\\
&&+g_{\rm B}(|\phi_7\rangle\langle\phi_9|+|\phi_8\rangle\langle\phi_{10}|)+\Omega_{\rm B}(t)(|\phi_9\rangle\langle\phi_{11}|+|\phi_{10}\rangle\langle\phi_{12}|)+\rm H.c.,
\end{eqnarray}
for which
\begin{eqnarray}\label{e3}
|\phi_1\rangle=|g\rangle_{\rm A}|g\rangle_{\rm B}|0\rangle_{c \rm A}|0\rangle_f|0\rangle_{c \rm B},\quad
|\phi_2\rangle=|e\rangle_{\rm A}|g\rangle_{\rm B}|0\rangle_{c \rm A}|0\rangle_f|0\rangle_{c \rm B},\nonumber\\
|\phi_3\rangle=|L\rangle_{\rm A}|g\rangle_{\rm B}|L\rangle_{c \rm A}|0\rangle_f|0\rangle_{c \rm B},\quad
|\phi_4\rangle=|R\rangle_{\rm A}|g\rangle_{\rm B}|R\rangle_{c \rm A}|0\rangle_f|0\rangle_{c \rm B},\nonumber\\
|\phi_5\rangle=|L\rangle_{\rm A}|g\rangle_{\rm B}|0\rangle_{c \rm A}|L\rangle_f|0\rangle_{c \rm B},\quad
|\phi_6\rangle=|R\rangle_{\rm A}|g\rangle_{\rm B}|0\rangle_{c \rm A}|R\rangle_f|0\rangle_{c \rm B},\nonumber\\
|\phi_7\rangle=|L\rangle_{\rm A}|g\rangle_{\rm B}|0\rangle_{c \rm A}|0\rangle_f|L\rangle_{c \rm B},\quad
|\phi_8\rangle=|R\rangle_{\rm A}|g\rangle_{\rm B}|0\rangle_{c \rm A}|0\rangle_f|R\rangle_{c \rm B},\nonumber\\
|\phi_9\rangle=|L\rangle_{\rm A}|e_L\rangle_{\rm B}|0\rangle_{c \rm A}|0\rangle_f|0\rangle_{c \rm B},\quad
|\phi_{10}\rangle=|R\rangle_{\rm A}|e_R\rangle_{\rm B}|0\rangle_{c \rm A}|0\rangle_f|0\rangle_{c \rm B},\nonumber\\
|\phi_{11}\rangle=|L\rangle_{\rm A}|L\rangle_{\rm B}|0\rangle_{c \rm A}|0\rangle_f|0\rangle_{c \rm B},\quad
|\phi_{12}\rangle=|R\rangle_{\rm A}|R\rangle_{\rm B}|0\rangle_{c \rm A}|0\rangle_f|0\rangle_{c \rm B}.
\end{eqnarray}
$|L(R)\rangle$ denotes a single left(right)-circularly polarized photon state. Now we set a set of orthogonal states
\begin{eqnarray}\label{e4}
|\psi_k\rangle=\frac{1}{\sqrt2}(|\phi_{2k+1}\rangle+|\phi_{2k+2}\rangle),\quad|\psi_k^-\rangle=\frac{1}{\sqrt2}(|\phi_{2k+1}\rangle-|\phi_{2k+2}\rangle),
\end{eqnarray}
with $k=1,2,3,4,5$. Then Hamiltonian~(\ref{e2}) becomes
\begin{eqnarray}\label{e5}
H(t)&=&\Omega_{\rm A}(t)|\phi_1\rangle\langle\phi_2|+\sqrt2g_{\rm A}|\phi_2\rangle\langle\psi_1|
+g_{\rm B}(|\psi_3\rangle\langle\psi_4|+|\psi_3^-\rangle\langle\psi_4^-|)\nonumber\\
&&+v(|\psi_1\rangle\langle\psi_2|+|\psi_1^-\rangle\langle\psi_2^-|+|\psi_2\rangle\langle\psi_3|+|\psi_2^-\rangle\langle\psi_3^-|)\nonumber\\
&&+\Omega_{\rm B}(t)(|\psi_4\rangle\langle\psi_5|+|\psi_4^-\rangle\langle\psi_5^-|)+\rm H.c..
\end{eqnarray}
Because $|\psi_k^-\rangle$ will not be involved during the whole evolution if $|\phi_1\rangle$ is the initial state, so Hamiltonian (\ref{e5}) becomes
\begin{eqnarray}\label{e6}
H(t)&=&\Omega_{\rm A}(t)|\phi_1\rangle\langle\phi_2|+\sqrt2g_{\rm A}|\phi_2\rangle\langle\psi_1|+v(|\psi_1\rangle\langle\psi_2|+|\psi_2\rangle\langle\psi_3|)\nonumber\\
&&+g_{\rm B}|\psi_3\rangle\langle\psi_4|+\Omega_{\rm B}(t)|\psi_4\rangle\langle\psi_5|+\rm H.c..
\end{eqnarray}

Next, for further simplification, we set $\sqrt2g_{\rm A}=v=g_{\rm B}=g$ and rewrite Hamiltonian~(\ref{e6}) as
\begin{eqnarray}\label{e7}
H(t)&=&H_0+V(t),\nonumber\\
H_0&=&g(|\Psi_1^+\rangle\langle\Psi_1^+|-|\Psi_1^-\rangle\langle\Psi_1^-|)+\sqrt3g(|\Psi_2^+\rangle\langle\Psi_2^+|-|\Psi_2^-\rangle\langle\Psi_2^-|),\nonumber\\
V(t)&=&\frac{\Omega_{\rm A}(t)}{\sqrt3}|\phi_1\rangle[\langle\Psi_d|-\frac{\sqrt3}{2}(\langle\Psi_1^+|+\langle\Psi_1^-|)
+\frac{1}{2}(\langle\Psi_2^+|+\langle\Psi_2^-|)]\nonumber\\
&&+\frac{\Omega_{\rm B}(t)}{\sqrt3}[|\Psi_d\rangle+\frac{\sqrt3}{2}(|\Psi_1^+\rangle+|\Psi_1^-\rangle)
+\frac{1}{2}(|\Psi_2^+\rangle+|\Psi_2^-\rangle)]\langle\psi_5|+\rm H.c.,
\end{eqnarray}
with the following transformations
\begin{eqnarray}\label{e8}
&|\Psi_d\rangle=\frac{1}{\sqrt3}(|\phi_2\rangle-|\psi_2\rangle+|\psi_4\rangle),\quad|\Psi_1^{\pm}\rangle=-\frac{1}{2}[|\phi_2\rangle\pm(|\psi_1\rangle-\psi_3\rangle)-|\psi_4\rangle],\nonumber\\
&|\Psi_2^{\pm}\rangle=\frac{1}{2\sqrt3}[|\phi_2\rangle+2|\psi_2\rangle\pm\sqrt3(|\psi_1\rangle+\psi_3\rangle)+|\psi_4\rangle].
\end{eqnarray}
Then after performing the unitary transformation $U=\exp(-iH_0t)$ and neglecting high oscillating terms under the limit condition $\Omega_{\rm A(B)}(t)\ll2g$, we simplify Hamiltonian~(\ref{e7}) to an effective Hamiltonian
\begin{eqnarray}\label{e9}
H_e(t)=\Omega_1(t)|\phi_1\rangle\langle\Psi_d|+\Omega_2(t)|\Psi_d\rangle\langle\psi_5|+\rm H.c.,
\end{eqnarray}
with $\Omega_1(t)=\Omega_{\rm A}(t)/\sqrt3$ and $\Omega_2(t)=\Omega_{\rm B}(t)/\sqrt3$.

\section{Superadiabatic scheme for fast generating two-atom 3D entanglement}

Instantaneous eigenstates of Hamiltonian~(\ref{e9}) with eigenvalues $\eta_{\pm}=\pm\Omega(t)$ and $\eta_0=0$, respectively, are
\begin{eqnarray}\label{e10}
|n_{\pm}(t)\rangle&=&\frac{1}{\sqrt{2}}[
\sin\theta_0(t)|\phi_{1}\rangle{\pm}|\Psi_{d}\rangle+\cos\theta_0(t)|\psi_{5}\rangle],\nonumber\\
|n_{0}(t)\rangle&=&\cos\theta_0(t)|\Psi_{1}\rangle-\sin\theta_0(t)|\Psi_{2}\rangle,
\end{eqnarray}
where $\Omega(t)=\sqrt{\Omega_1(t)^2+\Omega_2(t)^2}$ and $\tan\theta_0(t)=\Omega_1(t)/\Omega_2(t)$.
We transform $H_e(t)$ to the adiabatic frame by performing the unitary transformation $U_0(t)=\sum_{k=\pm,0}|n'_{k}\rangle\langle n_{k}(t)|$. At each instant in time, $U_0(t)$ maps the adiabatic eigenstate $n_{k}(t)$ onto the
time-independent state $|n'_{k}\rangle$. In the adiabatic frame, the Hamiltonian~(\ref{e9}) becomes
\begin{eqnarray}\label{e11}
H_1(t)&=&U_0(t)H_e(t)U_0^\dag(t)+i\dot{U}_0(t)U_0^\dag(t)\nonumber\\
&=&\Omega(t)[|n'_{+}\rangle\langle n'_{+}|-|n'_{-}\rangle\langle n'_{-}|]+\frac{\dot{\theta}_0(t)}{\sqrt2}[i|n'_{+}\rangle\langle n'_{0}|+i|n'_{-}\rangle\langle n'_{0}|+\rm H.c.].
\end{eqnarray}
The effective system evolution will adiabatically follow one of states $\{|n'_{0,\pm}\rangle\}$ with adiabatic approximation $|\dot{\theta}_0(t)|\ll\sqrt2\Omega(t)$ which needs very long runtime. For shortening runtime, Demirplack and Rice~\cite{DR} and Berry~\cite{MVB2009} proposed that adding a suitable counterdiabatic~(CD) Hamiltonian $H_{\rm CD}(t)$ to the original Hamiltonian can suppress transitions between different eigenstates. In the adiabatic frame CD Hamiltonian may be $-i\dot{U}_0(t)U_0^\dag(t)$, which is written in $\{|\phi_{1}\rangle,|\Psi_{d}\rangle,|\psi_{5}\rangle\}$ frame by
\begin{eqnarray}\label{e12}
H_{\rm CD}^{(1)}(t)=-iU_0^\dag(t)\dot{U}_0(t)=i \dot{\theta}_0(t)(|\phi_{1}\rangle\langle\psi_{5}|-|\psi_{5}\rangle\langle\phi_{1}|).
\end{eqnarray}
CD Hamiltonian~(\ref{e12}) needs a direct coupling between $|\phi_1\rangle$ and $|\psi_5\rangle$, which is too hard to implement in practice for such a complex system.

Superadiabatic states~(instantaneous eigenstates of $H_1(t)$) with eigenvalues $\eta'_{\pm}=\pm \Omega'(t)$ and $\eta'_0=0$, respectively, are
\begin{eqnarray}\label{e13}
|n''_{\pm}(t)\rangle&=&\frac{1}{{2}}\{
i[1\pm \cos\theta_1(t)]|n'_{+}\rangle
\pm\sqrt2\sin\theta_1(t)|n'_{0}\rangle+i[1\mp\cos\theta_1(t)]|n'_{-}\rangle\},\nonumber\\
|n''_0(t)\rangle&=&\frac{1}{\sqrt{2}}[
-i\sin\theta_1(t)|n'_{+}\rangle+
\sqrt2\cos\theta_1(t)|n'_{0}\rangle+i\sin\theta_1(t)|n'_{-}\rangle],
\end{eqnarray}
for which $\Omega'(t)=\sqrt{\dot{\theta}_0(t)^2+\Omega(t)^2}$ and $\tan\theta_1(t)=\dot{\theta}_0(t)/\Omega(t)$. Then we transform $H_1(t)$ to the superadiabatic frame by the unitary transformation $U_1(t)=\sum_{k=\pm,0}|\widetilde {n}_{k}\rangle\langle n''_{k}(t)|$. Analogous to the adiabatic CD Hamiltonian~(\ref{e12}), the superadiabatic CD Hamiltonian $-i\dot{U}_1(t)U_1^\dag(t)$ is written in $\{|\phi_{1}\rangle,|\Psi_{d}\rangle,|\psi_{5}\rangle\}$ frame by
\begin{eqnarray}\label{e14}
H_{\rm CD}^{(2)}(t)&=&-iU_0^\dag(t)U_1^\dag(t)\dot{U}_1(t)U_0(t)\nonumber\\
&=&\dot{\theta}_1(t)[-\cos\theta_0(t)|\phi_{1}\rangle\langle\Psi_{d}|
+\sin\theta_0(t)|\Psi_{d}\rangle\langle\psi_{5}|]+\rm H.c..
\end{eqnarray}
$H_{\rm CD}^{(2)}(t)$ is satisfactory because it has the same form as the effective Hamiltonian~(\ref{e9}).

We regard $\Omega'_1(t)=-\dot{\theta}_1(t)\cos\theta_0(t)$ and $\Omega'_2(t)=\dot{\theta}_1(t)\sin\theta_0(t)$ as two auxiliary pulses added to the pulses $\Omega_1(t)$ and $\Omega_2(t)$, respectively. Then modified pulses $\Omega''_1(t)=\Omega_1(t)+\Omega'_1(t)$ and $\Omega''_2(t)=\Omega_2(t)+\Omega'_2(t)$ can drive the effective system to evolve along one of the superadiabatic state in equation~(\ref{e13}). Therefore, if related parameters meet time boundary conditions $\{\theta_1(0)=\theta_1(t_f)=0,\theta_0(0)=0,\theta_0(t_f)=-\arctan\sqrt2\}$~($t_f$ is the final time), $|n''_0(t)\rangle$ will act as a medium state for achieving the expected transformation $|\phi_1\rangle\rightarrow\frac{1}{\sqrt3}(|\phi_{1}\rangle+\sqrt2|\psi_{5}\rangle)$. By this way, we obtain the 3D entanglement between two atoms in the superadiabatic scheme
\begin{eqnarray}\label{e15}
|\Psi_{\rm 3D}\rangle&=&\frac{1}{\sqrt3}(|\phi_{1}\rangle+|\phi_{11}\rangle+|\phi_{12}\rangle)\nonumber\\
&=&\frac{1}{\sqrt3}(|g\rangle_{\rm A}|g\rangle_{\rm B}+|L\rangle_{\rm A}|L\rangle_{\rm B}+|R\rangle_{\rm A}|R\rangle_{\rm B})\otimes|0\rangle_{c \rm A}|0\rangle_f|0\rangle_{c \rm B}.
\end{eqnarray}

\section{Numerical simulations}

\begin{figure}[htb]\centering
\centering
\includegraphics[width=\linewidth]{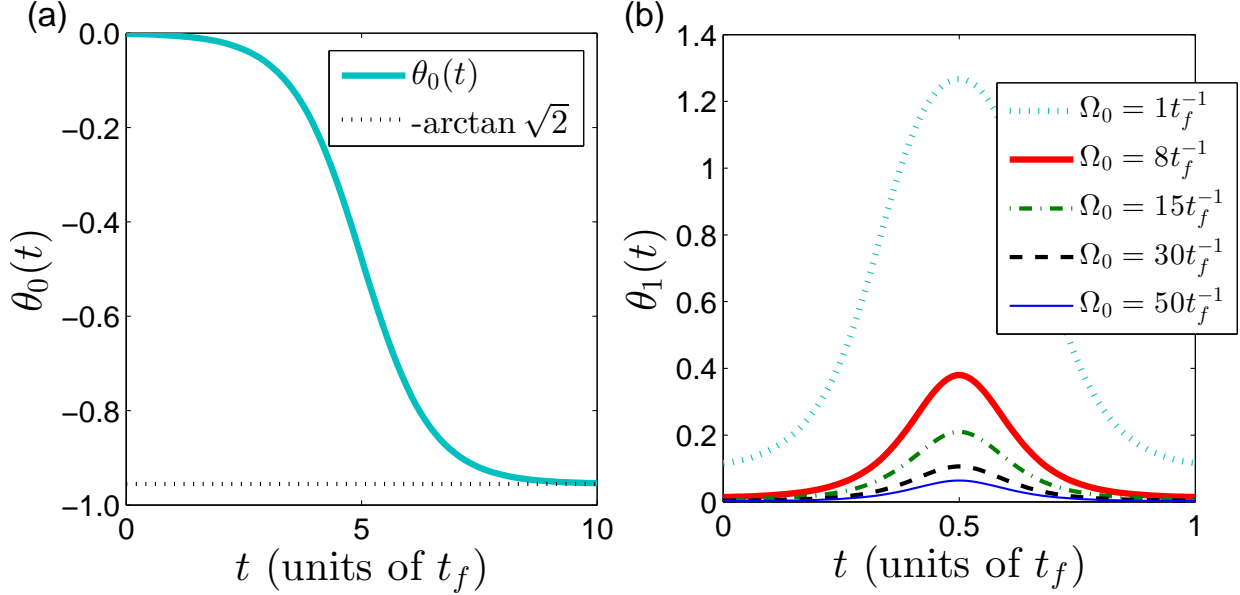}
\caption{(a)~Time dependence of $\theta_0(t)$ with an arbitrary $\Omega_0$; (b)~Time dependence of $\theta_1(t)$ with different values of $\Omega_0$.}\label{f2}
\end{figure}

\begin{figure}[htb]\centering
\centering
\includegraphics[width=\linewidth]{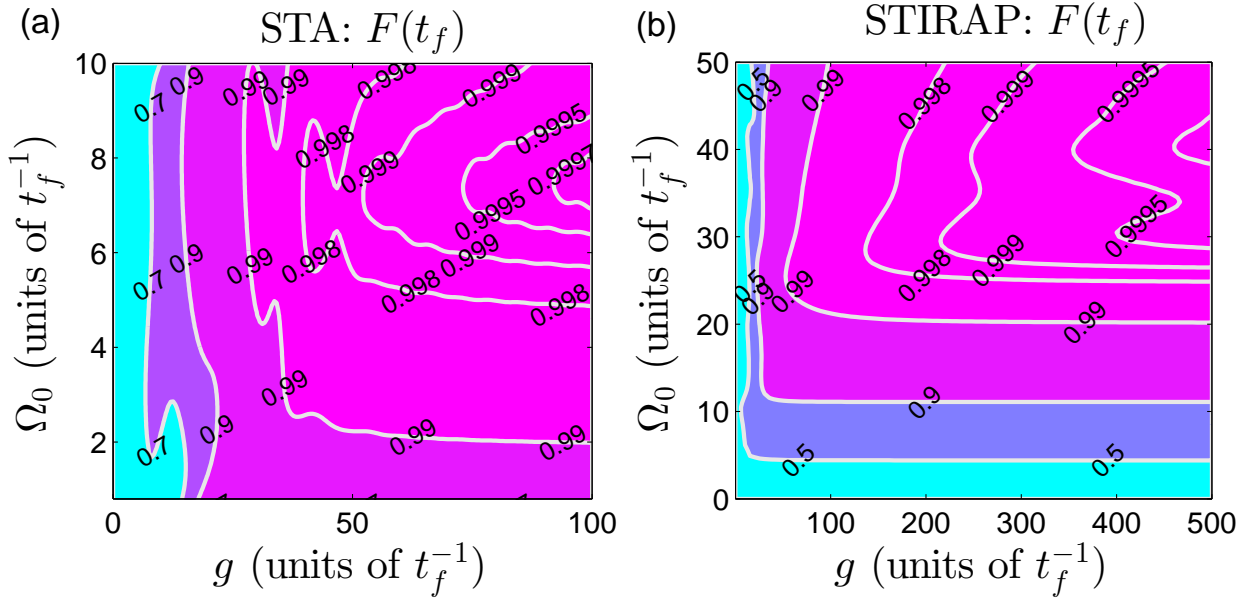}
\caption{Contour images for final fidelities versus $g$ and $\Omega_0$ in (a)~the STA scheme, i.e., superadiabatic scheme and (b)~the STIRAP scheme.}\label{f3}
\end{figure}

For meeting $\{\theta_1(0)=\theta_1(t_f)=0,\theta_0(0)=0,\theta_0(t_f)=-\arctan\sqrt2\}$, we choose $\Omega_2(t)$ and $\Omega_1(t)$ as~\cite{NTB2001}
\begin{eqnarray}\label{e16}
\Omega_2(t)&=&\frac{1}{\sqrt3}\Omega_0\exp[-(t-t_f/2-t_0)^2/t_c^2]+\Omega_0\exp[-(t-t_f/2+t_0)^2/t_c^2],\nonumber\\
\Omega_1(t)&=&-\frac{\sqrt2}{\sqrt3}\Omega_0\exp[-(t-t_f/2-t_0)^2/t_c^2],
\end{eqnarray}
with two related Gaussian parameters $t_0=0.18t_f$ and $t_c=0.24t_f$. In figure~\ref{f2}, we plot the time dependence of $\theta_0(t)$ and $\theta_1(t)$ determined by $\Omega_1(t)$ and $\Omega_2(t)$. $\{\theta_0(0)=0,\theta_0(t_f)=-\arctan\sqrt2\}$ can be always satisfied well with an arbitrary value of $\Omega_0$, while $\theta_1(0)=\theta_1(t_f)=0$ can be satisfied well only with a large enough value of $\Omega_0$. In order to choose suitable parameters, in figure~\ref{f3}(a), we plot the final fidelity $F(t_f)=|\langle\Psi_{\rm 3D}|\Psi(t_f)\rangle|^2$ versus $g$ and $\Omega_0$ in the superadiabatic scheme, in which $|\Psi(t_f)\rangle$ is the final state of the whole system governed by Hamiltonian (\ref{e1}) with Rabi frequencies $\Omega_{\rm A}(t)=\sqrt3\Omega''_1(t)$ and $\Omega_{\rm B}(t)=\sqrt3\Omega''_2(t)$ . As a contrast, in figure~\ref{f3}(b), we plot the final fidelity in the STIRAP scheme with Rabi frequencies $\Omega_{\rm A}(t)=\sqrt3\Omega_1(t)$ and $\Omega_{\rm B}(t)=\sqrt3\Omega_2(t)$. By converting the relation $g\sim t_f^{-1}$ into $t_f\sim g^{-1}$, we easily find that, for the same final fidelity, the operation time of the superadiabatic scheme is reduced to about $1/5$ of that of the STIRAP scheme, which proves that the scheme we proposed is fast indeed.
Besides, both from figures \ref{f3}(a) and \ref{f3}(b), we also see that the limit condition $\Omega_{\rm A(B)}(t)\ll2g$ is acting. Therefore, we adopt a pair of parameters $\Omega_0=8t_f^{-1}$ and $g=70t_f^{-1}$ for the superadiabatic scheme in following discussions.

\begin{figure}[htb]\centering
\centering
\includegraphics[width=\linewidth]{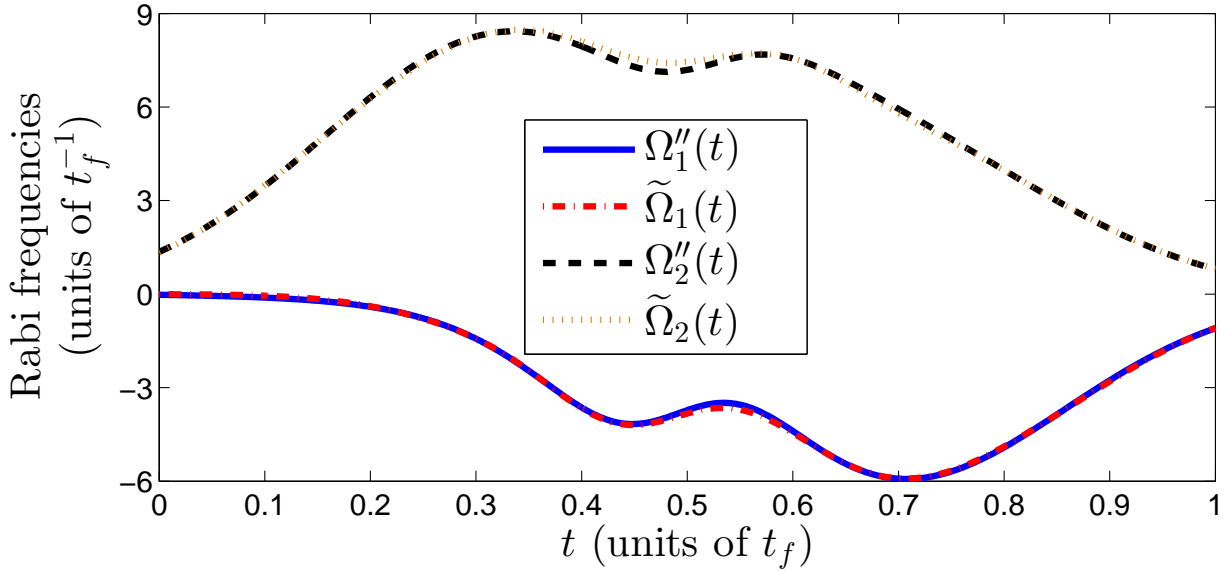}
\caption{Time dependence of $\Omega''_1(t)$, $\Omega''_2(t)$, $\widetilde{\Omega}_1(t)$ and $\widetilde{\Omega}_2(t)$.}\label{f4}
\end{figure}
\begin{figure}[htb]\centering
\centering
\includegraphics[width=\linewidth]{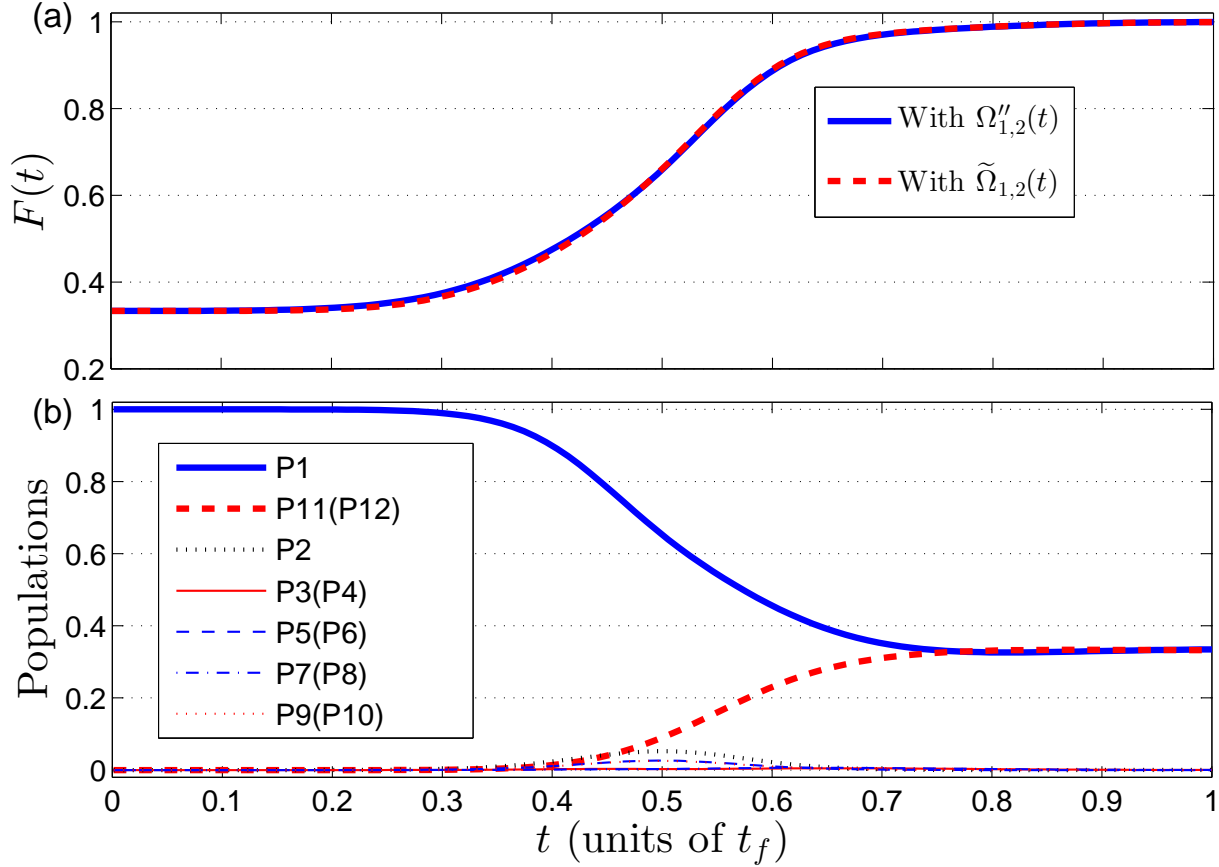}
\caption{(a)~Comparison between the fidelities for adopting \{$\Omega''_1(t)$,$\Omega''_2(t)$\} and \{$\widetilde{\Omega}_1(t)$,$\widetilde{\Omega}_2(t)$\}; (b)~Time evolutions of the populations for states in equation~(\ref{e3}), respectively, with \{$\widetilde{\Omega}_1(t)$,$\widetilde{\Omega}_2(t)$\}\}.}\label{f5}
\end{figure}
Since analytic functions of $\Omega''_1(t)$ and $\Omega''_2(t)$ are too complicated, for the experimental feasibility, we seek two superpositions of Gaussian functions by curve fitting to replace them, respectively
\begin{eqnarray}\label{e17}
\widetilde{\Omega}_1(t)=-\sum_{m=0}^4\Omega_{1m}\exp[-(t-\tau_{1m})^2/\chi_{1m}^2],\quad
\widetilde{\Omega}_2(t)=\sum_{m=0}^4\Omega_{2m}\exp[-(t-\tau_{2m})^2/\chi_{2m}^2],
\end{eqnarray}
with related parameters $\{
\Omega_{11}=1.4695/t_f,\Omega_{12}=2.4114/t_f,\Omega_{13}=1.9854/t_f,\Omega_{14}=4.4491/t_f,\tau_{11}=0.3733t_f,\tau_{12}=0.4424t_f,
\tau_{13}=0.6547t_f,\tau_{14}=0.7568t_f,\chi_{11}=0.1494t_f,\chi_{12}=0.0939t_f,\chi_{13}=0.1358t_f,\chi_{14}=0.2044t_f\}$ for $\widetilde{\Omega}_1(t)$ and $\{
\Omega_{21}=6.7888/t_f,\Omega_{22}=1.1904/t_f,\Omega_{23}=1.649/t_f,\Omega_{24}=5.4413/t_f,\tau_{21}=0.2814t_f,\tau_{22}=0.3712t_f,
\tau_{23}=0.5752t_f,\tau_{24}=0.6588t_f,\chi_{21}=0.2204t_f,\chi_{22}=0.12t_f,\chi_{23}=0.0987t_f,\chi_{24}=0.2475t_f\}$ for $\widetilde{\Omega}_2(t)$.
Through plotting figure~\ref{f4}, we see that the curve for $\widetilde{\Omega}_1(t)$~($\widetilde{\Omega}_2(t)$) is very close to that for $\Omega''_1(t)$~($\Omega''_2(t)$). In the following, for showing the effectiveness of two alternative Rabi frequencies, in figure~\ref{f5}(a) we plot time dependence of the fidelity for adopting $\Omega''_1(t)$ and $\Omega''_2(t)$ or $\widetilde{\Omega}_1(t)$ and $\widetilde{\Omega}_2(t)$. Highly approximate coincidences of two pairs of curves indicate that the alternative Rabi frequencies are pretty valid. For a further illustration, with $\widetilde{\Omega}_1(t)$ and $\widetilde{\Omega}_2(t)$, in figure~\ref{f5}(b) we plot time evolutions of the populations for all states in equation~(\ref{e3}), respectively, and the results show that the desired two-atom 3D entanglement $|\Psi_{\rm 3D}\rangle$ can be obtained near perfectly at $t=t_f$. What's more, we also see that the states not involved in $|\Psi_{\rm 3D}\rangle$ are hardly populated.

\begin{figure}[htb]\centering
\centering
\includegraphics[width=\linewidth]{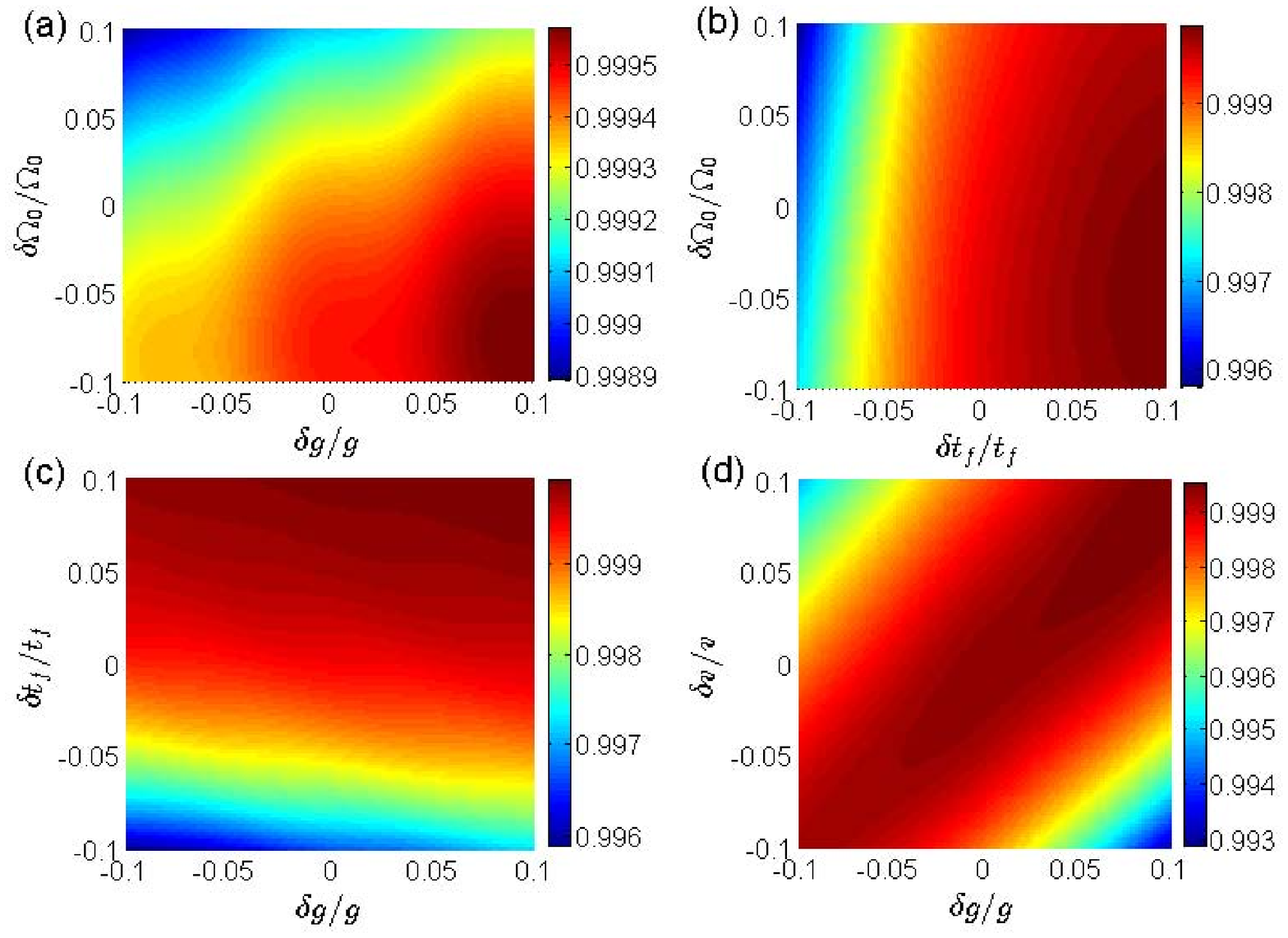}
\caption{Effects of variations in various parameters on the final fidelity.}\label{f6}
\end{figure}
Since most parameters are impossible to control perfectly in experiment, we should investigate the robustness of the scheme against variations in control parameters. Here we define $\delta x=x'-x$ as the deviation of $x$, in which $x$ denotes the ideal value and $x'$ denotes the actual value. In figure~\ref{f6}, we consider effects of variations in parameters involved in the superadiabatic scheme on the final fidelity for fast generating the 3D entanglement between two atoms. As we can see from figure~\ref{f6}, the final fidelity always keep over $0.993$ even when variations in two of parameters we consider are both up to  $|\delta x|=0.1x$, which indicates the superadiabatic scheme for fast generating the two-atom 3D entanglement is extremely robust against variations in control parameters. By the way, figure~\ref{f6}(d) also shows the condition $v=g$ is a little bit critical for the high-fidelity generation of the target state.

Finally, taking decoherence caused by atomic spontaneous emissions and photon leakages from the cavity-fiber system into account, the evolution of the whole system will be dominated by the master equation
\begin{eqnarray}\label{e18}
\dot{\rho}(t)&=&-i[H(t),\rho(t)]\nonumber\\
&&-\sum_{j=g,L,R}\frac{\gamma_{j}^{\rm A}}{2}[\sigma^{\rm A}_{e,e}\rho(t)-2\sigma^{\rm A}_{j,e}\rho(t)\sigma^{\rm A}_{e,j}+\rho(t)\sigma^{\rm A}_{e,e}]
\nonumber\\
&&-\sum_{j=0,L,R}\sum_{i=L,R}\frac{\gamma_{j,i}^{\rm B}}{2}[\sigma^{\rm B}_{e_i,e_i}\rho(t)-2\sigma^{\rm B}_{j,e_i}\rho(t)\sigma^{\rm B}_{e_i,j}+\rho(t)\sigma^{\rm B}_{e_i,e_i}],\nonumber\\
&&-\sum_{i=L,R}\sum_{l={\rm A,B}}\frac{\kappa_{i}^{l}}{2}[a_{li}^{\dag}a_{li}\rho(t)-2a_{li}\rho(t)a_{li}^{\dag}+\rho(t)a_{li}^\dag
a_{li} ]\nonumber\\
&&-\sum_{i=L,R}\frac{\kappa_{i}^{f}}{2}[b_{i}^{\dag}b_{i}\rho(t)-2b_{i}\rho(t)b_{i}^{\dag}+\rho(t)b_{i}^\dag b_{i}],
\end{eqnarray}
where $H(t)$ is Hamiltonian~(\ref{e1}); $\gamma_{j}^{l}$ is the photon leakage rate of atom $l$ from excited states to the ground state $|j\rangle$; $\kappa_{i}^{l}$ is the $i$-circular polarized photon leakage rate of cavity $l$; $\kappa_{i}^{f}$ is the $i$-circular polarized photon leakage rate of the fiber; $\sigma_{m,n}=|m\rangle\langle n|$. For simplicity, we assume $\gamma_{j}^{\rm A}=\gamma_{j,i}^{\rm B}=\gamma/2$, $\kappa_{i}^{l}=\kappa_{i}^{f}=\kappa$.

\begin{figure}[htb]\centering
\centering
\includegraphics[width=\linewidth]{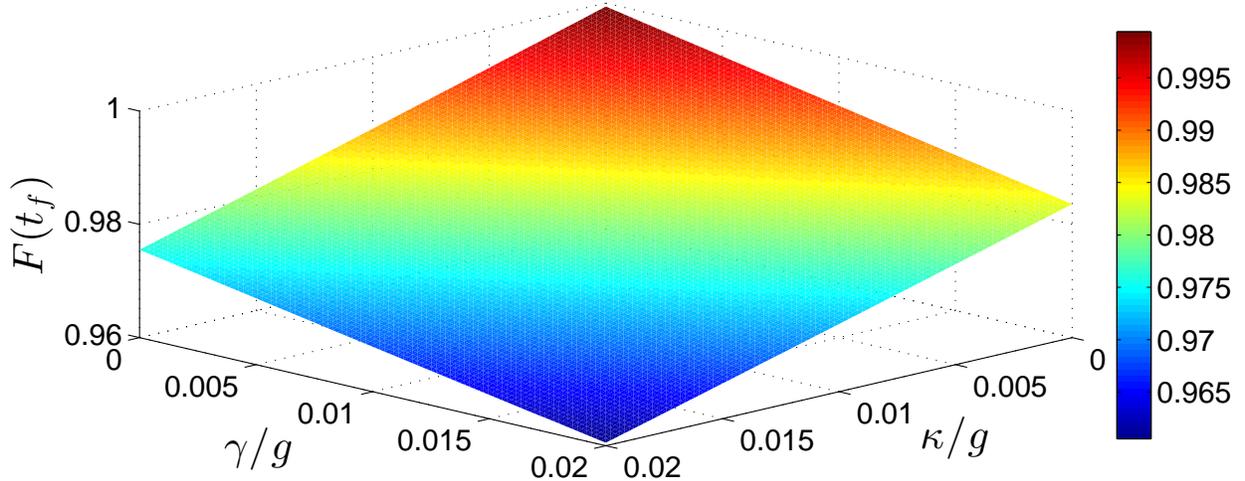}
\caption{Final fidelity versus $\kappa/g$ and $\gamma/g$.}\label{f7}
\end{figure}
Based on the master equation, in figure~(\ref{f7}) we plot the final fidelity $F(t_f)=|\langle\Psi_{\rm 3D}|\rho(t)|\Psi_{\rm 3D}\rangle|$ for fast generating two-atom 3D entanglement versus $\kappa/g$ and $\gamma/g$. We can clearly see that the superadiabatic scheme is very robust against decoherence induced by atomic spontaneous emissions and cavity-fiber photon leakages, because even when $\kappa=\gamma=0.02g$, the final fidelity can keep very high $F(t_f)>0.96$. For a real experiment, $^{87}\emph{Rb}$ and a set of predicted cavity-QED parameters $(\lambda, \kappa, \gamma)/2\pi=(750, 3.5, 2.62)$ MHz~\cite{STO2003,STK2005} can be used for generating two-atom 3D entanglement in the superadiabatic scheme, and the final fidelity can reach $99.93\%$.

\section{Conclusion}
In conclusion, we have implemented the fast generation of 3D entanglement between two atoms in an experimentally feasible superadiabatic scheme. Under the certain limit condition, the complicated system is simplified to a three-state system, which makes the superadiabatic scheme more convenient to be applied for generating two-atom 3D entanglement. Because of the compensation for non-adiabatic couplings, the adiabatic approximation is not needed. Compared with TQD, the superadiabatic scheme is more complicated just in terms of mathematical calculations. But it does not require a direct coupling between the initial and finial states, which greatly enhances experimental feasibility. In addition, the results of numerical simulations show that the superadiabatic scheme is strongly robust against variations in various parameters and decoherence caused by atomic spontaneous emissions and cavity-fiber photon leakages.

\begin{center}
{\bf{ACKNOWLEDGMENT}}
\end{center}
This work was supported by the National Natural Science Foundation of China under Grants No. 11464046.

\end{document}